\documentclass[proceedings]{JHEP3} % 10pt is ignored!
\PrHEP{ hep2003}

\usepackage{epsfig,multicol}			% please use epsfig.

\newbox\mybox
\newcommand\fverb{\setbox\mybox=\hbox\bgroup\verb}
\newcommand\fverbdo{\egroup\medskip\noindent\fbox{\unhbox\mybox}\ }
\newcommand\fverbit{\egroup\item[\fbox{\unhbox\mybox}]}

\title{Signature~of~HDM~clustering~at~{\sc Planck}~angular~scales}

\author{\speaker{L.A. Popa}
\footnote{Further address: Institute for Space Sciences, 
Bucharest-Magurele R-76900, Romania} ,
                 C. Burigana, and N. Mandolesi \\
	IASF/CNR, Istituto di Astrofisica Spaziale e Fisica Cosmica, Sezione di Bologna,
        Consiglio Nazionale delle Ricerche, Via Gobetti 101, Bologna I-40129, Italy\\
	E-mail: \email{popa burigana mandolesi @bo.iasf.cnr.it}}

\conference{AHEP-2003}

\abstract{We present the CMB anisotropy
induced by the non-linear perturbations in
the massive neutrino density
associated to the non-linear gravitational clustering.
We show  that the non-linear time varying  potential
induced by the gravitational clustering process
generates metric perturbations
that affect the time evolution of the density fluctuations in all the
components of the expanding Universe,
leaving imprints on the CMB
anisotropy power spectrum at subdegree angular scales.
For a neutrino fraction in agreement with that
indicated by the astroparticle and nuclear physics experiments
and a cosmological accreting mass comparable
with the mass of known  clusters,
we find that CMB anisotropy measurements with
{\sc Planck} angular resolution and sensitivity
possibly combined to other precise cosmological 
observations will allow the detection of the dynamical, linear and non-linear
effects of the neutrino gravitational clustering.}

\begin{document}

\section{Introduction}

The atmospheric neutrino results from the  Super-Kamiokande \cite{sk}
and MACRO \cite{mac} experiments indicate that neutrinos oscillate,
these data being consistent with $\nu_{\mu} \leftrightarrow \nu_{\tau}$
oscillations. The small value of the difference of the squared masses
($5 \times 10^{-4}{\rm eV}^2  \leq \Delta m^{2}
\leq 6 \times 10^{-3}{\rm eV}^2$)
and the strong mixing angle ($ \sin^{2} 2\theta \geq 0.82$) suggest that
these neutrinos are nearly equal in mass
as predicted by many models of particle physics beyond the
standard model.
Also, the LSND experiment \cite{Ath98} support
$\nu_{\mu}\leftrightarrow \nu_{e}$ oscillations
($\Delta m^{2} \leq 0.2$eV$^{2}$) and
other different types of solar neutrino experiments
\cite{Bah98} suggest that $\nu_e$ could oscillate to a sterile
neutrino $\nu_e \leftrightarrow \nu_s$
($\Delta m^{2} \simeq 10^{-5}$eV$^2$).
The direct implication of neutrino oscillations is the existence of
a non-zero neutrino mass in the eV range, and consequently
a not negligible hot dark matter (HDM) contribution $\Omega_{\nu} \neq 0$
to  the total energy density of the Universe.

We study the cosmic microwave background (CMB) secondary anisotropies
induced by the non-linear perturbations in the massive neutrino density
associated to the non-linear gravitational clustering.
The extent to which the massive neutrinos can cluster gravitationally
depends on their mass and the parameters
of the fiducial cosmological model
describing
the present Universe.\\
The linear perturbation theory describes
accurately the growth of density fluctuations from the early Universe
until a redshift $z \sim 100$.
The solution involves the integration of
coupled and linearized Boltzmann, Einstein and fluid equations
\cite{mb95} that describes the time evolution of the metric
perturbations in the perturbed density field and the time evolution of the
density fields in the perturbed space-time for all the relevant
species (e.g., photons, baryons, cold dark matter, massless
and massive neutrinos). At lower redshifts  the gravitational
clustering becomes a non-linear process
and the solution  relies on numerical simulations. \\
Through numerical simulations, we compute the CMB
anisotropy in the non-linear
stages of the evolution of the
Universe when clusters and superclusters of galaxies start to form
producing a non-linear gravitational potential varying with time.
By using a standard particle-mesh method
we analyze the imprint of the dynamics of the
neutrino gravitational clustering on the CMB anisotropy
power spectrum in a  flat $\Lambda$CHDM model
with
neutrino fractions
$f_{\nu}=\Omega_{\nu}/(\Omega_b+\Omega_{cdm})$=0.06, 0.11, 0.16
corresponding to
$\Omega_{\nu}$=0.022 ($m_\nu$=0.78eV), 0.037 ($m_\nu$=1.35 eV),
0.053 ($m_\nu$=1.89eV), assuming three massive neutrino flavors.
This model is consistent with the LSS data
and the WMAP anisotropy latest measurements \cite{sper03, ver03}
allowing in the same time a pattern of neutrino masses
consistent with the results from neutrino oscillation
and double beta decay experiments.

\section{The neutrino gravitational infall}

In the expanding Universe, neutrinos decouple from the other
species when the ratio of their interaction rate to the expansion rate
falls below unity. For neutrinos with masses in the
eV range the decoupling temperature is T$_{D} \sim$ 1MeV, occurring
at a redsfit z$_{D} \sim $10$^{10}$ \cite{fkt83}.
At this time neutrinos behave like relativistic particles with
a  pure Fermi-Dirac
phase-space distribution:
\begin{eqnarray}
f_{\nu}(q,a)=\frac{1} {e^{E_{\nu}/T_{\nu} +1}},
\hspace{0.3cm} E_{\nu}=\sqrt {q^2+a^2 m^2_{\nu}} \, ,
\end{eqnarray}
where ${\vec q}$ is the neutrino comoving momentum,
${\vec q}=a {\vec p}$, ${\vec p}$ being
the neutrino 3-vector momentum, $E_{\nu}$ is the energy of
neutrino with mass $m_{\nu}$ and $a = 1/(1+z)$ is the cosmic scale factor
evolving with the time, $t$ ($a_0$=1 today).\\
As  neutrinos are collisionless particles, they
can significantly interact with  photons, baryons and cold dark matter
particles only via gravity.
The neutrino phase space density
is constrained by the Tremaine \& Gunn
criterion \cite{tg79} 
that put  limits on the neutrino energy density inside
the gravitationally bounded objects:
in the cosmological models
involving a HDM component (the CHDM models)
the compression fraction of neutrinos through a cluster
$f(r)=\rho_{\nu}/\rho_{cdm}$ (where $r$ is the cluster radius)
never exceeds the background
ratio $\Omega_{\nu}/\Omega_{cdm}$ \cite{kof85}.
Because the formation of
galaxies and clusters is a dynamical time process,
the differences introduced in the gravitational
potential due to neutrino gravitational clustering
generate metric perturbations that
affect the evolution  of the density
fluctuations of all the components of the expanding Universe.
Fig.~1 presents the  evolution of the
projected mass distributions of cold dark matter plus baryons
and neutrinos obtained from numerical simulations at
few redshift values $z$.
One can see that neutrinos are accreted by the cold dark
matter and baryons, contributing in dynamic way to the
gravitational clustering process.\\
\begin{figure}
\epsfig{figure=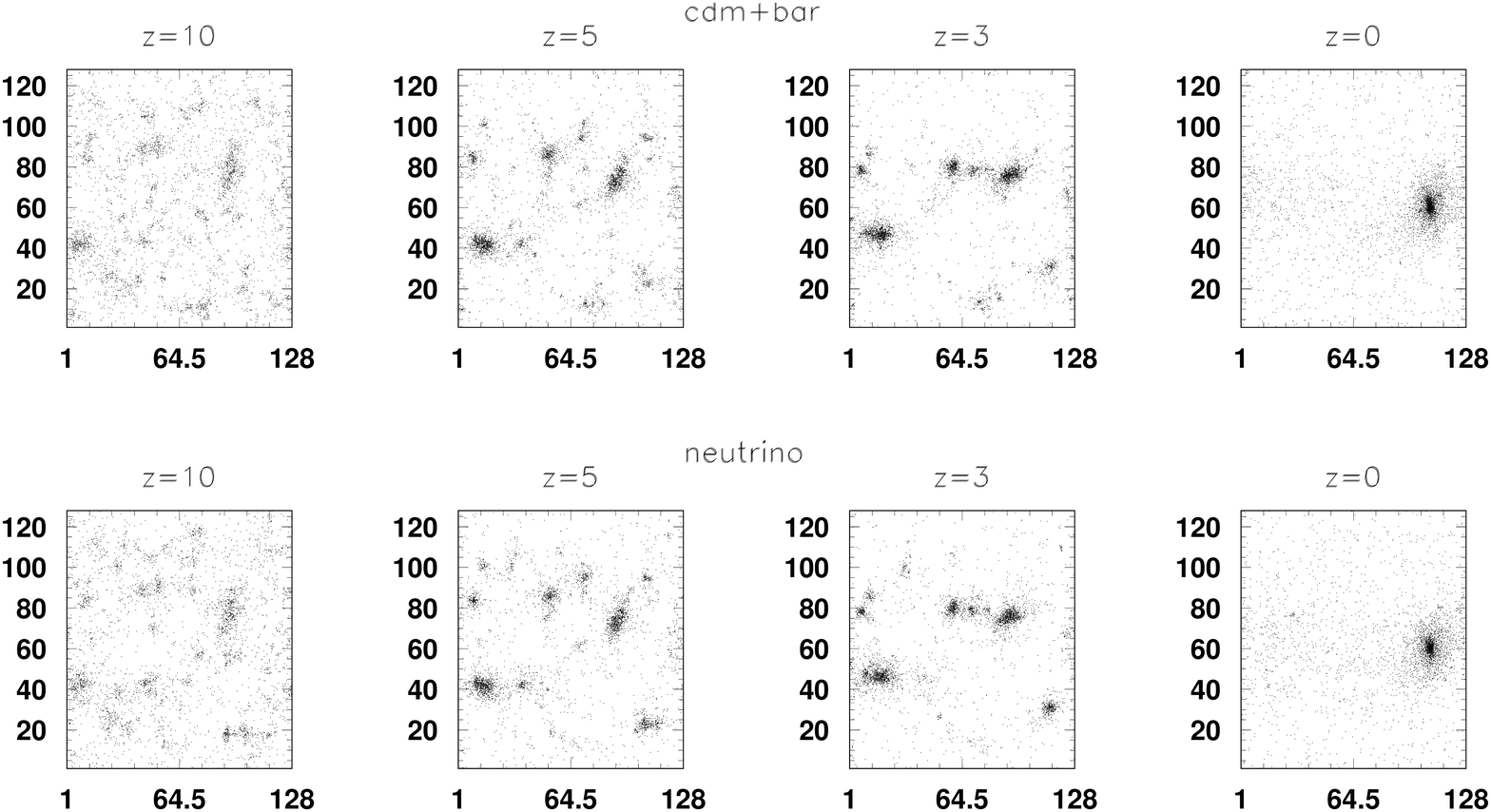,width=16cm,height=10cm}
\caption{The evolution with the redshift of the projected mass distributions of
cold dark matter plus baryons
(upper row) and neutrinos (lower row) obtained from numerical simulation
of $128^3$ cold dark matter plus baryons (the total mass of $8 \times 10^{16} M_{\odot}$)
and $10 \times 128^3$ neutrinos
(the total mass of $4.8 \times 10^{18} M_{\odot}$) in a
box of size 128 Mpc, for the $\Lambda$CHDM model with the neutrino
fraction
$f_{\nu}=0.06$ ($\sum_i m_{\nu_{i}}\approx 0.7$ eV). 
[Units of axes are in Mpc].}
\end{figure}
Neutrinos cannot cluster via gravitational instability
on distances below the free-streaming distance $R_{fs}$
\cite{bes80, bs83, ma00}.
The neutrino free-streaming distance is related to the
causal comoving
horizon distance $\eta(a)$ through \cite{dgs96}:
\begin{eqnarray}
R_{fs}(a)=\frac{1}{k_{fs}}=\frac{\eta(a)}{\sqrt{1+(a/a_{nr})^2}}
\,{\rm Mpc},
\hspace{0.35cm}
\eta(a)=\int^a_0 \frac{da}{a^2 H(a)},
\end{eqnarray}
where $a_{nr}$ is the value of the scale factor
when massive neutrinos start to become non-relativistic
($a_{nr}=(1+z_{nr})^{-1} \approx 3k_{B}T_{\nu,0}/ m_{\nu}c^2$)
and $H(a)$ is the Hubble expansion rate:
\begin{eqnarray}
H^2(a)= \left({da/dt \over a}\right)^2 =
\frac{8 \pi G}{3}[\Omega_m/a^3+\Omega_r/a^4+
\Omega_{\Lambda}+\Omega_k/a^3].
\end{eqnarray}
In the above equation $G$ is the gravitational constant,
$\Omega_m=\Omega_b+\Omega_{cdm}+\Omega_{\nu}$
is the matter energy density parameter, $\Omega_b$, $\Omega_{cdm}$,
$\Omega_{\nu}$ being the energy density parameters of baryons, cold dark
matter and massive neutrinos, $\Omega_{r}$ is the
radiation energy density parameter
that includes the contribution from photons and relativistic neutrinos ,
$\Omega_{\Lambda}$ is the vacuum (or cosmological constant)
energy density parameter,
$\Omega_k=1-\Omega_m-\Omega_{\Lambda}$ is the energy density parameter
related to the curvature of the Universe.\\
\begin{figure}
\epsfig{figure=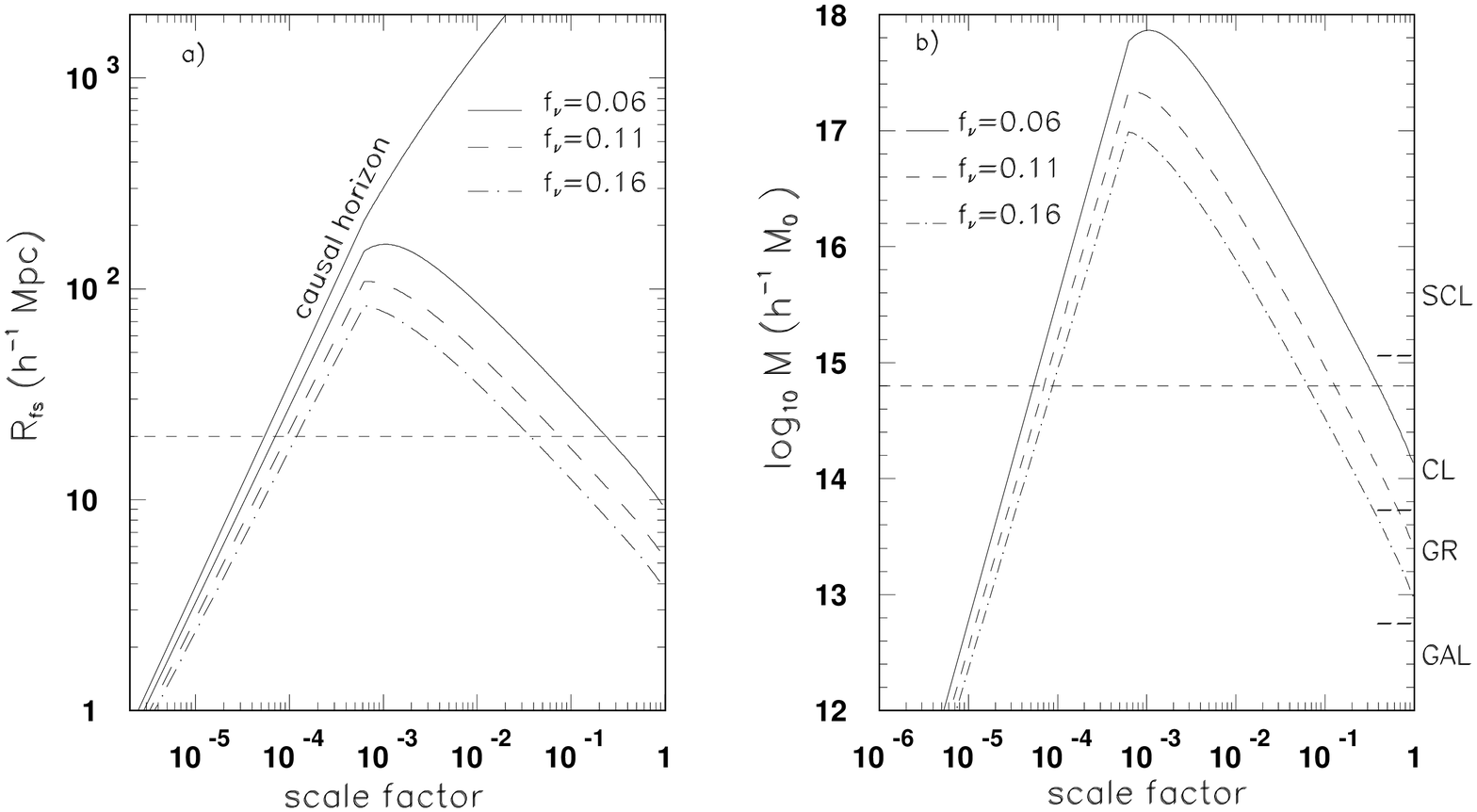,height=6cm,width=15cm}
\caption{Panel a): dependence of the
neutrino free-streaming distance $R_{fs}$ on the scale factor.
We show also a specific scale, $\lambda$=20$h^{-1}$Mpc, constant
in comoving coordinates (horizontal dashed line)
and the evolution
with the scale factor of the causal horizon distance
for our cosmological models.
Panel b): dependence on the scale factor of the
mass $M(R_{fs})$ of the perturbation at the scale $R_{fs}$.
We show also the mass of the perturbation at the scale $\lambda$
(horizontal dashed line) and indicate
the typical mass ranges for galaxies (GAL),
groups (GR), clusters (CL) and superclusters (SCL). }
\end{figure}
$R_{fs}$ defines the minimum linear dimension that a
neutrino perturbation should have in order to survive
the free-streaming.
In the spherical approximation,
the minimum comoving mass of a
perturbation that should contain clusterized neutrinos,
corresponds to \cite{kt90}
%\begin{eqnarray}
$$M(R_{fs}) =\frac{\pi}{6} R_{fs}^3 \rho_m
 \approx 1.5 \times 10^{11}\,(\Omega_m h^2)
(R_{fs}/ {\rm Mpc})^3 h^{-1}M_{\odot} \, ,$$ 
%\nonumber
%\end{eqnarray}
where $\Omega_m$ is the matter energy density parameter.\\
We show in Fig.~2
the dependence of the causal horizon distance $\eta(a)$,
 the neutrino free-streaming distance $R_{fs}$,
[panel a)] and of the mass
$M(R_{fs})$ [panel b)] on the cosmic scale factor.
The cosmological model is the $\Lambda$CHDM model
with different neutrino fractions $f_{\nu}$.
One can see that at early times, when neutrinos  are relativistic,
the free-streaming distance is approximately the causal horizon distance.
After neutrinos become non-relativistic ($a_{nr}\sim 10^{-4}$ for
our cosmological models) the free-streaming distance
decreases with time, becoming smaller
than the causal horizon distance.
The time behaviors of $R_{fs}$ and $M(R_{fs})$
show that neutrino can cluster gravitationally on increasingly
smaller scales at latter times.
If the causal horizon $\eta(a)$ is
large enough to encompass the wavelength $\lambda$,
the neutrino gravitational infall perturbs the
growth of the perturbations for this mode,
leaving  imprints in the CMB angular power spectrum.
Perturbations on scales $\lambda < R_{fs}$ ($k > k_{fs}$)
are damped due to the neutrino free-streaming while
the perturbations on scales $\lambda > R_{fs}$ ($k < k_{fs}$)
are affected only by gravity.
In the Newtonian limit, the neutrino gravitational clustering
can be described as a deviation
from the background by a potential $\Phi$
given by the Poisson equation:
\begin{equation}
{\nabla}^2 \Phi({\vec r},a) =4 \pi G a^2
\rho_m(a)\delta_m({\vec r},a) \, ,
\end{equation}
where ${\vec r}$ is the position 3-vector,
$\rho_m(a)$ is the
matter density
and $\delta_m({\vec r},a)$ is the matter density
fluctuation;
$\delta_m=(\rho_b \delta_b + \rho_{cdm} \delta_{cdm}
+ \rho_{\nu} \delta_{\nu})/ \rho_m$, where
$\rho_{cdm}$, $\rho_b$, $\rho_{\nu}$ and
$\delta_{cdm}$, $\delta_b$, $\delta_{\nu}$
are the density and density fluctuations
of cold dark matter particles, baryons and neutrinos. \\
The equations governing the motion of each particle species
(cold dark matter plus baryons and neutrinos)
in the expanding Universe
are given by \cite{gb94}:
\begin{eqnarray}
\frac{{d \vec q}}{da}= -a\,H(a) \, {\vec \nabla} \Phi ,
\hspace{0.5cm}  \frac{ d{\vec r} } { da }
={\vec q}\, (a^3 \, H(a))^{-1},
\end{eqnarray}
where ${\vec q}$ is
the comoving momentum and $H(a)$ is given by the equation (2.3).\\
The Newtonian description given by the equations (2.4) and (2.5)
applies in the limit of the week gravitational field if, at each time step,
the size of the non-linear structures
is much smaller than the causal horizon
size (the background curvature is negligible).

The cosmological models involving
massive neutrinos show a characteristic
scale dependence of the perturbation growth rates
\cite{ma96, he98, pbm01}.

We evolve the system of baryons plus cold dark matter particles
and neutrinos according to the equation (2.5)
for the non-linear scales involved in the computation
of the CMB anisotropy (0.06Mpc$^{-1} \leq k \leq $0.52Mpc$^{-1}$),
starting from the beginning of the non-linear regime
of cold dark matter plus baryons component.\\
The initial positions and velocities of neutrinos
and baryons plus cold dark
matter particles can be generated at each
spatial wave number $k$
from the corresponding matter density fluctuations
power spectra at the
present time by using
the Zel'dovich approximation \cite{zel70}.
The matter power spectra was normalized on the basis of the
analysis of the local cluster X-ray
temperature function \cite{ecf96}.
We performed simulations with $128^3$ cold dark matter plus
baryon particles and
$10 \times 128^3$ neutrinos.
The neutrinos and the baryons plus cold dark matter particles
was randomly placed
on $128^3$ grids, with  comoving spacing $r_0$ of 0.5~h$^{-1}$Mpc.
The high number of neutrinos and this comoving spacing
ensure a precision high enough for a correct sampling of the neutrino
phase space distribution
\cite{pbm01}. \\
According to the Zel'dovich approximation, the perturbed comoving
position of each particle ${\vec r}( {\vec r_0}, a)$ and its peculiar
velocity ${\vec v}( {\vec r_0},a)$ are related to the
fluctuations of the density field
$\delta \rho({\vec r_0},a,k)$ through:
\begin{eqnarray}
{\vec r}( {\vec r_0},k,a) &=& {\vec r_0}+D(k,a)
{\vec d}({\vec r_0}) \, , \;\;\;
{\vec v}( {\vec r_0},k,a) = {\dot D}(k,a){\vec d}({\vec r_0}) \, , \\
{\vec \nabla}{\vec d}({\vec r_0})
&=& D^{-1}(k,a) \delta \rho({\vec r_0},k,a) \nonumber \, ,
\end{eqnarray}
where ${\vec r_0}$ is the coordinate corresponding to the
unperturbed comoving position,
${\vec d}({\vec r_0})$ is the displacement field
and $D(k,a)$ is the growth function of perturbations
corresponding to each cosmological model.\\
At each wave number $k$ used in the computation of the CMB anisotropy
we compute the perturbed particles
comoving positions and  peculiar velocities
at the beginning of the non-linear regime $a_{nl}$, by using
the set of equations (2.4)--(2.5).
\begin{figure}
\vspace{-4.cm}
\epsfig{figure=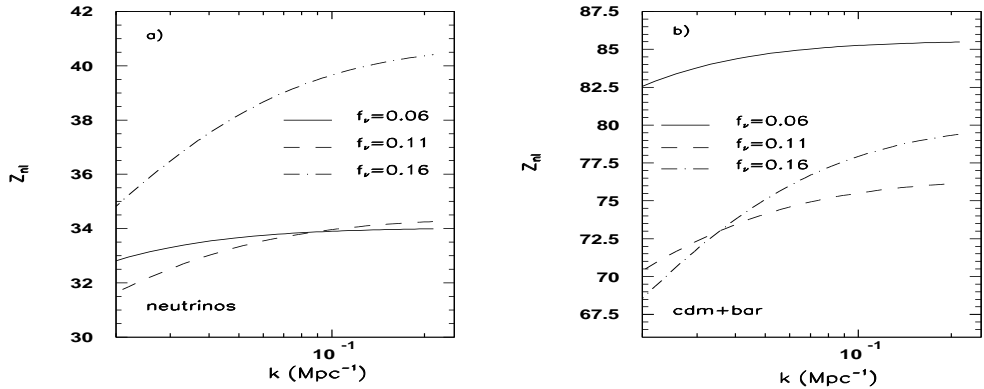,height=10cm,width=15cm}
\caption{The scale dependence of $z_{nl}$
%=1/a_{nl}-1$
on $k$ for neutrinos (panel a) and baryons plus cold
dark matter particles (panel b).}
\end{figure}
We assign to each particle a momentum
according to the growth  function,
when the power of each mode is randomly selected
from a Gaussian distribution with the mean accordingly
to the  corresponding power spectrum
\cite{hr91, gh93, bert95}.
In the computation of the set of equations (2.6) we consider
only the growing modes, the non-linear power spectra up to
$k_{max}=6.28$~h~Mpc$^{-1}$,
and neglect the contribution of the redshift distortions.\\
We show in Fig.~3 the dependence on the spatial wave number $k$
of the redshift $z_{nl}=1/a_{nl}-1$
%scale factor $a_{nl}$
for each component.
One can see from Fig.~3 that
neutrinos (panel~a)  enter in the non-linear regime
later than cold dark matter particles and baryons (panel~b).
Thus, the neutrino halo of the cluster starts to form after
the cold dark matter plus baryon halo is advanced
in the non-linear stage,
causing the accretion of neutrinos from
the background.\\
At each spatial wavenumber $k$ we evolve the
particles positions and velocities according to
the set of equations~(2.4)--(2.5).
We start this process  from the scale factor $a^{cdm}_{nl}$
at which cold dark matter particles plus baryons start
to enter in the non-linear regime.
At each time step, the density on the
mesh is obtained from the particle positions using
the Cloud-in-Cell method and
equations (2.6) are solved by using 7-point discrete analog
of the Laplacian operator and the FFT technique.
 The particle positions and
velocities are then advanced in time with a time step $da$
required by the computation of the CMB anisotropy
power spectra.
The system of particles was evolved until the
scale factor $a_{st}$
when
it reaches its virial \cite{peac01} equilibrium.
%The virial theorem states that
%a system of particles evolving in the gravitational field
%achieves its equilibrium  state when
%the time averaged kinetic energy $E_k$ and the potential energy $E_{p}$
%of the system are related through: $E_{p}=-2E_{k}$.

\section{Imprints of neutrino gravitational
clustering at {\sc Planck} angular scales}

As the  anisotropy produced by the non-linear
density perturbations depends on the
time variations of the
spatial gradients of the gravitational
potential produced by different components
(cold dark matter, baryons, neutrinos),
we calculate the CMB anisotropy in the presence of the
gravitational clustering by using N-body simulation
in large boxes with the side of $128$ Mpc,
that include all non-linear scales used in the computation of
the CMB anysotropy power spectrum from
$\lambda_{min}\approx 12$Mpc ($k_{max}\approx 0.52$Mpc$^{-1}$)
to $\lambda_{max}\approx$ 110 Mpc ($k_{min}\approx 0.06$Mpc$^{-1}$),
taking into account the time evolution of all non-linear
density perturbations influencing the CMB power spectrum
(see also Fig.~1). One should note that $\lambda_{max}$ corresponds to
the comoving horizon size at the matter-radiation equality
for our cosmological models
($\lambda_{eq} \approx 16 \,\Omega_m^{-1}h^{-2}$Mpc).
The non-linear structures are assumed to be formed by
two components: cold dark matter plus baryons and neutrinos in the form
of three massive neutrino flavors,
both components evolving in the gravitational field created
by themself. For the purpose of this work we neglect the
hydrodinamical effects \cite{qis95}. 
%This approach is justified as
%the contributions to the CMB anisotropy of the hot baryonic gas
%is proved to be negligible \cite{qis95}.
The neutrino gravitational clustering
can affect both the homogeneous and the inhomogeneous components
of the gravitational field.
The changes in the  homogeneous component
of the gravitational field
are determined by the
changes of the energy density of neutrinos and cold dark matter particles
plus baryons. They affect the Hubble expansion rate,
the sound horizon distance and the neutrino
free-streaming distance.
The changes in the  inhomogeneous component
of the gravitational field are determined by the
changes in the energy density for all matter components
and the changes in the neutrino phase space distribution function.
They affect
the growth of the energy density perturbations of
cold dark matter, baryons, photons, massive and massless
neutrinos.
\begin{figure}
\vspace{-6cm}
\epsfig{figure=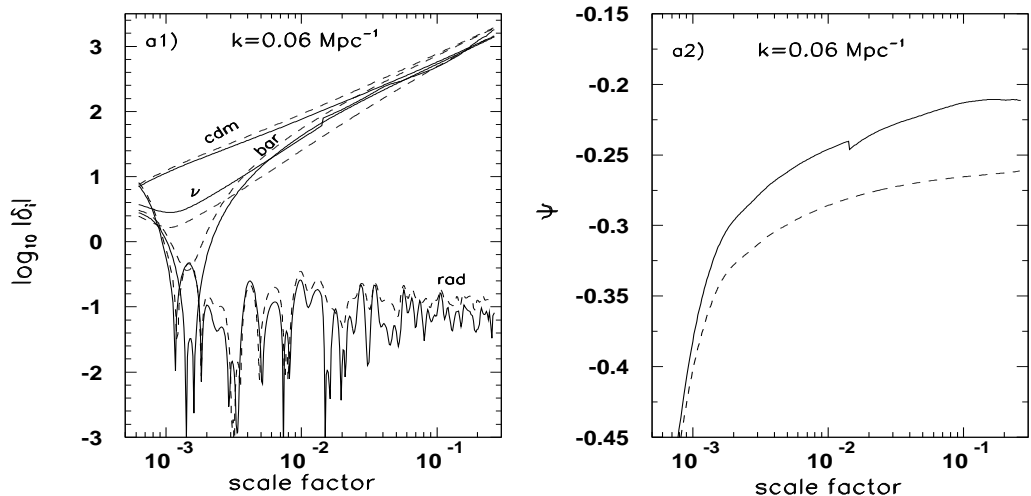,height=13cm,width=15cm}
\caption{Panel a1): time evolution of the energy density
perturbations of the different components as computed by including
linear and non-linear effects of neutrino gravitational clustering (solid
lines) and by neglecting the non-linear aspects
of neutrino gravitational clustering (dashed lines):
cold dark matter (cdm), baryons (bar), massive neutrinos ($\nu$),
and massless neutrinos plus photons (rad).
Panel a2): the same as in panel a), but for
the time evolution of the gravitational field
[k=0.06Mpc$^{-1}$ and $f_{\nu}=0.06$].}
\end{figure}
Panel a1) of Fig.~4 presents the evolution
with the scale factor
of the energy density perturbations
of different components in the non-linear regime,
for the mode $k=0.06$ Mpc$^{-1}$ (solid lines). For comparison,
we plot also (dashed lines) the energy density perturbations
of the different components
obtained for the same mode $k$ by neglecting
non-linear aspects of neutrino gravitational clustering (linear regime).
Panel a2) of Fig.~4 presents the evolution
with the scale factor of the scalar potential $\Psi$
of the conformal Newtonian gauge line element, that plays
the role of the gravitational potential in the Newtonian
limit  \cite{bar80, ks84},
by including (solid line) or not (dashed line)
the non-linear effects of neutrino gravitational clustering
(for the transformation relation between
the scalar potentials of the synchronous gauge
and conformal Newtonian gauge
see equation (18) from \cite{mb95}).
As we have shown before, the difference in the evolution
of a perturbation mode $k$ depends on how this mode
relates to the neutrino free-streaming wave number $k_{fs}$.
Considering that our simulation at each time step
is a sample of the evolution of
the matter in the non-linear regime,
we study the imprint of the gravitational clustering
on the CMB anisotropy power spectrum
by smoothing the density field
obtained from simulation at each time step
with a filter with the scale $R_{fs}$ corresponding to
the cluster mass value $M(R_{fs})$. For each non-linear mode
$k$ only the perturbations with the mass $M \leq M(R_{fs})$
are  taken into account for the computation of
the CMB anisotropy power spectrum.
\begin{figure}

\vspace{-5cm}
\epsfig{figure=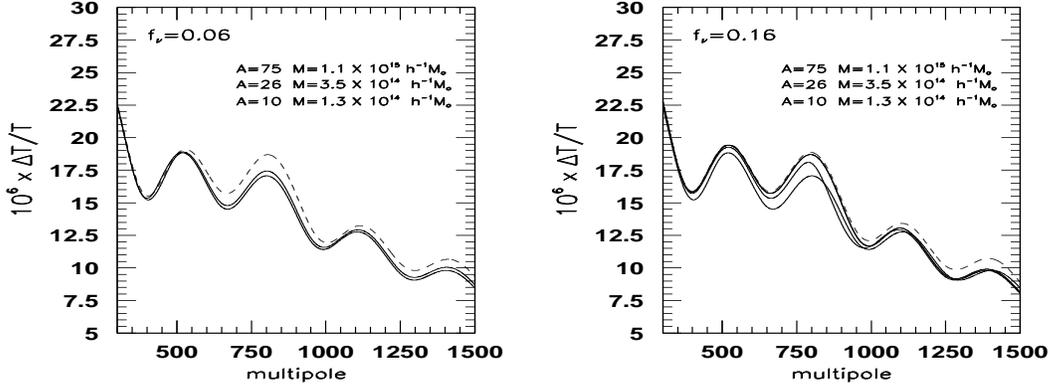,width=15cm,height=10cm}
\caption{The  imprint of the neutrino
linear and non-linear gravitational clustering on the CMB
anisotropy power spectrum expressed in terms of $\Delta T/T$
obtained for the filtering perturbation with the mass $M$.
We report in the panels the richness
of these perturbations from the bottom to the top 
according to the increasing of the power at multipoles 
about $\ell \sim 750$. 
In each panel, the dashed line corresponds to  the fiducial
$\Lambda$CHDM cosmological model, without including the non-linear effects
of neutrino gravitational clustering.}
\end{figure}
Fig.~5 presents some of our computed CMB anisotropy power spectra
obtained when different filtering mass values $M(R_{fs})$
are considered.
It is usual to use the Coma cluster as the mass normalization point
($M_{{\rm Coma}}=1.45 \times 10^{15}$h$^{-1}$M$_{\odot}$);
for the Coma cluster we assume a richness
${\cal A}_{{\rm Coma}}$=106.
According to \cite{kas98},
%Kashlinsky,
the relation between the
mass of the perturbation and the richness ${\cal A}$
of the corresponding cluster can be written in the form
%\begin{eqnarray}
$$M=M_{{\rm Coma}} \frac{{\cal A}}{{\cal A}_{{\rm Coma}}}=
1.45 \times 10^{15}\left( \frac{{\cal A}}{106}
\right) h^{-1}M_{\odot} \, .$$ 
%\nonumber
%\end{eqnarray}
By comparing the angular power spectra obtained
including or not the non-linear effects of neutrino gravitational
clustering, we find a decrease of the CMB angular
power spectrum induced by the neutrino non-linear gravitational
clustering of $\Delta T/T \approx 10^{-6}$
for angular resolutions between $\sim 4$ and 20 arcminutes, depending
on the cluster mass and neutrino fraction $f_{\nu}$.

Clearly, new high sensitivity and resolution
anisotropy experiments will have the
capability to detect the neutrino gravitational clustering effect.
In particular, the instruments on-board the ESA {\sc Planck} 
satellite~\footnote{http://astro.estec.esa.nl/Planck} 
will measure the CMB angular power spectrum
with very high sensitivity up to multipoles $\ell \sim 1000-2000$
with a stringent control of the systematic effects.
Fig.~6 compares {\sc Planck} and WMAP~\footnote{http://lambda.gsfc.nasa.gov} 
performances. The CMB angular  
power spectrum is reported without beam smoothing and by taking into
account the beam window functions of several {\sc Planck} frequency 
channels and of the highest WMAP frequency channel (which is very close to
that of the LFI 70~GHz channel). The corresponding angular power spectra
of the residual nominal white noise (i.e. after the subtraction of 
the expection of its angular power spectrum) are also displayed. 
Of course, binning the power spectrum on a suitable range of multipoles,
as usual at high $\ell$, will allow to recover the CMB power spectrum
also at multipoles higher that those corresponding to the crossings
between the noise and CMB power spectra reported in the 
figure~\footnote{
At $\ell$ higher than $\simeq 1500-2000$ the 
confusion noise from extragalactic source fluctuations   
dominates over the instrumental noise, while Galactic 
foregrounds are relevant at multipoles less than few hundreds.}.

%\vspace{-5cm}
%\epsfig{figure=f8.eps,width=15cm,height=15cm}
\begin{figure}
\epsfig{figure=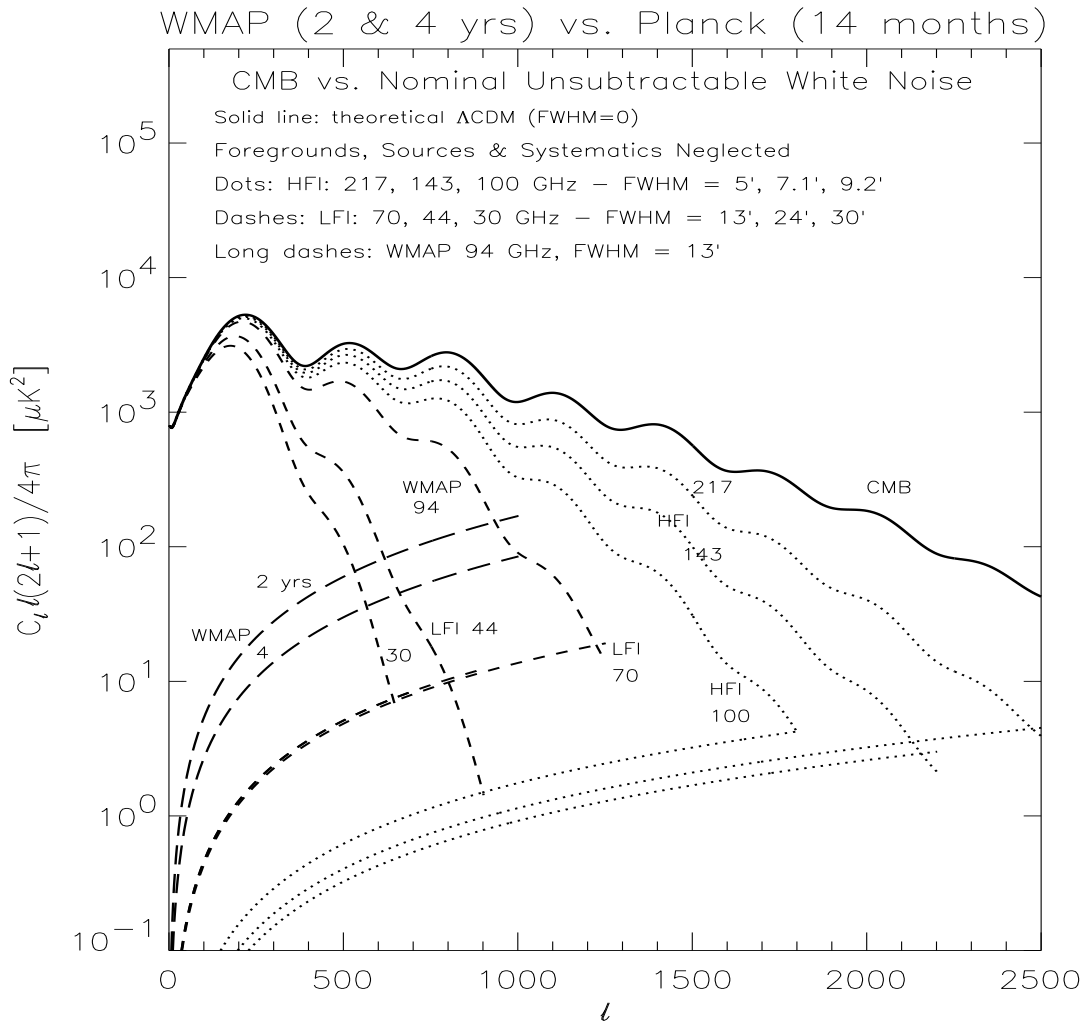,width=15cm,height=14cm}
\caption{Comparison between {\sc Planck} and WMAP resolution and 
sensitivity. For each considered frequency
channel, the crossing between the CMB convolved angular power spectrum
and the unsubtractable instrumental white noise angular power spectrum
indicates the multipole value where the signal to noise ratio 
($\ell$ by $\ell$) is close to unity.}
\end{figure}

The characteristic angular scale left by the
the neutrino gravitational clustering on the CMB
anisotropy power spectrum is given by
%\begin{equation}
$$\theta=\frac{R_{fs}}{\eta_0-\eta(a)} \, ,$$
%\end{equation}
where $R_{fs}$ is the scale of the filtering perturbation
with the mass $M(R_{fs})$,
$\eta(a)$ is the particle horizon distance
at the time at which the non-linear perturbation mode
$k$ cross the horizon
and $\eta_0$ is the particle horizon at the present time.
Fig.~7 (left panel) presents the evolution of the
characteristic scale $\theta$ and of the corresponding
multipole order of the CMB anisotropy power spectrum  with
the mass $M(R_{fs})$.

\begin{figure}[t]
\centering
\begin{tabular}{cc}
   \includegraphics[width=7cm,height=7cm]{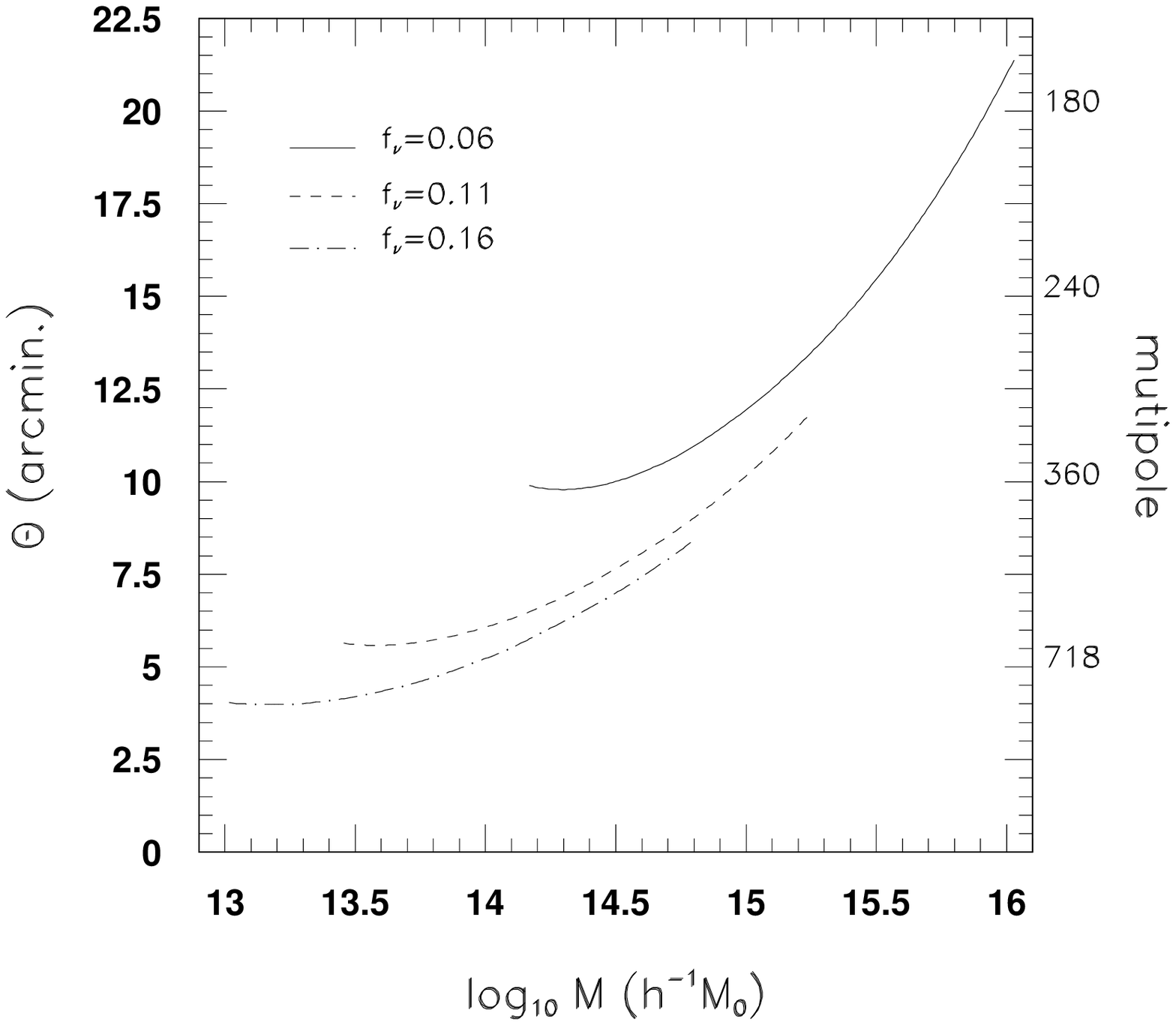}&
   \includegraphics[width=7cm,height=7cm]{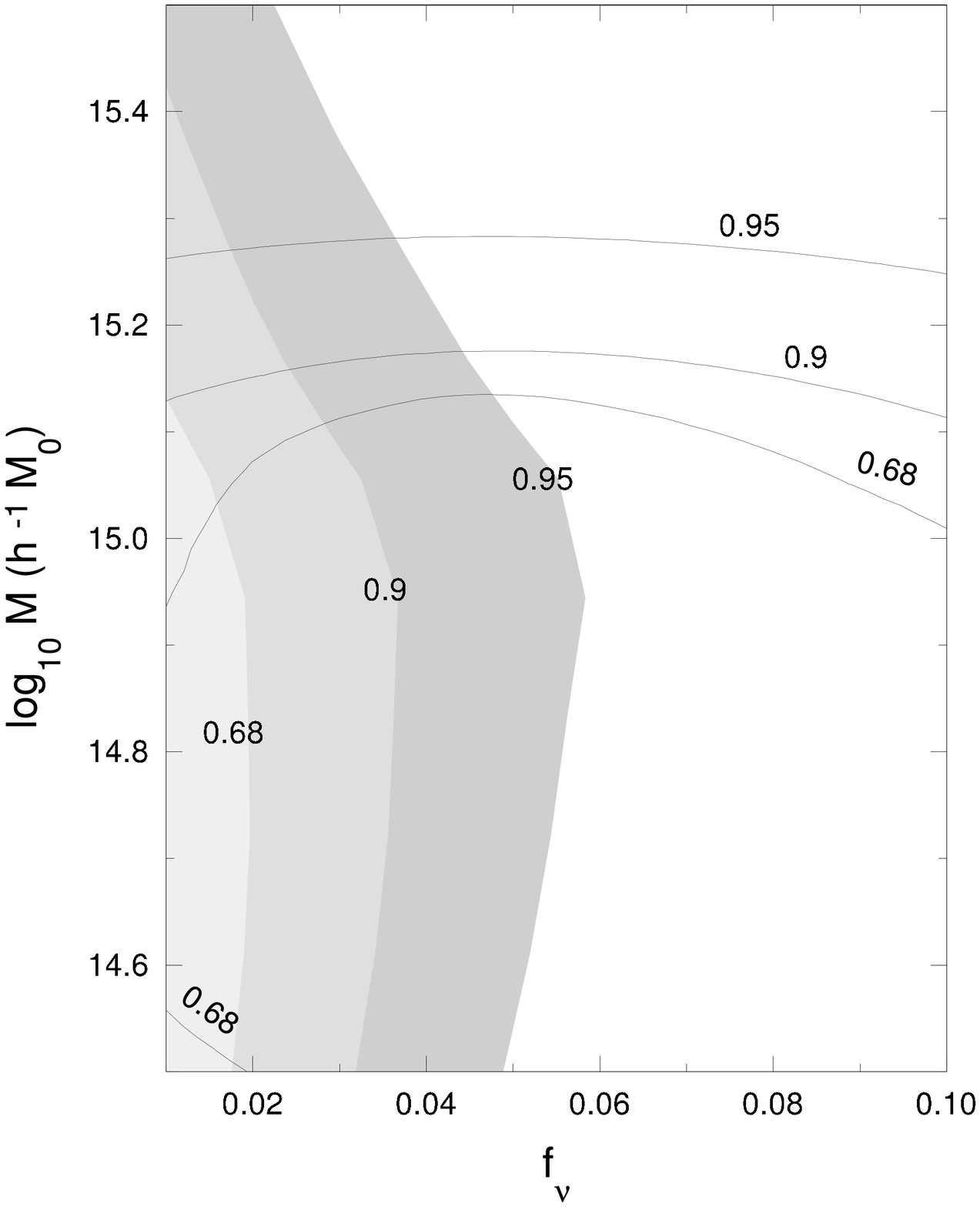}
\end{tabular}
\caption{Left panel: evolution of the characteristic angular scale 
$\theta$ of the neutrino gravitational clustering
with $M(R_{fs})$ for different neutrino fractions.
We report also the corresponding multipole orders
$l\sim \theta^{-1}$ of the CMB anisotropy power spectrum.
Right panel: confidence regions of the $f_{\nu} - M$ parameter space
as potentially detectable by {\sc Planck} by using the CMB
anisotropy measurements in the presence of the gravitational clustering
assuming known the other main cosmological parameters (grey regions) or
by jointly recover $f_{\nu}$, $M$ and the other main cosmological
parameters (solid lines).}
\end{figure}
%%%
Fig.~7 (right panel) presents few confidence regions
of the $f_{\nu} - M$ parameter space that can be potentially detected by
the {\sc Planck} surveyor by using the CMB anisotropy measurements
in the presence of the gravitational clustering, under the hypothesis
that the other cosmological parameter can be measured with other
observations (grey regions) and by using the {\sc Planck} data to
jointly determine $f_{\nu}$, $M$ and the other cosmological parameters 
(solid lines).
We consider for this computation only the {\sc Planck} ``cosmological''
channel between 70 and 217~GHz, a sky coverage $f_{sky}=0.8$
and neglect for simplicity the
foreground contamination \cite{pbm01}. By assuming
known the other main cosmological parameters,
we obtain a neutrino
fraction $f_{\nu} \approx 0.011 \pm 0.007$ for an accreting mass
$M \approx (8.2 \pm 3.1) \times 10^{14} h^{-1}M_{\odot}$
(errors at 68\% confidence level).
By assuming
known the other main cosmological parameters,
we obtain a neutrino
fraction $f_{\nu} \approx 0.011 \pm 0.007$ for an accreting mass
$M \approx (8.2 \pm 3.1) \times 10^{14} h^{-1}M_{\odot}$
(errors at 68\% confidence level).
{\sc Planck} surveyor will have
in principle the capability
to measure the non-linear imprints of the neutrino gravitational
clustering on the CMB anisotropy power spectrum for a neutrino
mass range in agreement with that
indicated by the astroparticle and nuclear physics experiments
and a cosmological accreting mass comparable
with the mass of the known clusters.
Of course, even with the high sensitivity and resolution
of {\sc Planck} it is hard to firmly constrain $f_{\nu}$ and $M$
by jointly recovering the other cosmological
parameters, a goal that can be achieved in combination
with other precise cosmological information, such as
galaxy large scale structure surveys, measures of element abundances
from big-bang nucleosynthesis and Type 1a supernovae observations.

\section{Conclusions}

We study the CMB anisotropy
induced by the non-linear perturbations in the massive neutrino density
associated to the non-linear gravitational clustering.
Through numerical simulations, we compute the CMB anisotropy
angular power spectrum in the non-linear stages of the evolution of the 
Universe
when clusters and superclusters start to form,
producing a non-linear time varying gravitational potential.\\
We found that the non-linear time varying  potential
induced by the gravitational clustering process
generates metric perturbations
that affect the time evolution of the density fluctuations in all the
components of the expanding Universe, leaving imprints on the CMB
anisotropy power spectrum at subdegree angular scales.
The magnitude of the induced anisotropy
and the characteristic angular scale depends on how each non-linear mode
$k$ of the perturbations relates to the neutrino free-streaming wavenumber
$k_{fs}$ at each evolution time step.
By smoothing the density field obtained from simulations
with a filter with the scale corresponding to the
cluster scale,
we find an imprint on the CMB
anisotropy power spectrum of amplitude
$\Delta T/T \approx 10^{-6}$
for angular resolutions between $\sim 4$ and 20 arcminutes, depending
on the cluster mass and neutrino fraction $f_{\nu}$.\\
This result suggests that the CMB anisotropy experiments
with such levels of sensitivities and angular resolutions
should detect the dynamical effect of
the non-linear gravitational clustering.
For a neutrino fraction in agreement with that
indicated by the astroparticle and nuclear physics experiments
and a cosmological accreting mass comparable
with the mass of known the clusters,
we find that CMB anisotropy measurements with
{\sc Planck} angular resolution and sensitivity
in combination with other precise cosmological observations
will allow the detection of the dynamical, linear and non-linear,
effects of the neutrino gravitational clustering.

\acknowledgments
LAP  acknowledge  the financial support from the European
Space Agency.

\end{document}